\documentclass[pra,aps,groupedaddress,twocolumn,floatfix,nofootinbib]{revtex4}
\usepackage[utf8]{inputenc} 
\usepackage{graphicx}
\usepackage{soul}

\newcommand{\beq}{\begin{equation}}
\newcommand{\eeq}{\end{equation}}
\newcommand{\beqa}{\begin{eqnarray}}
\newcommand{\eeqa}{\end{eqnarray}}

\def\opone{\leavevmode\hbox{\small1\normalsize\kern-.33em1}}

\begin{document}

\title{Classical and intuitionistic mathematical languages shape our understanding of time in physics}
\author{Nicolas Gisin \\
\it \small   Group of Applied Physics, University of Geneva, 1211 Geneva 4,    Switzerland}

\date{\small \today}
\begin{abstract}
Physics is formulated in terms of timeless classical mathematics. A formulation on the basis of intuitionist mathematics, built on time-evolving processes, would offer a perspective that is closer to our experience of physical reality.
\end{abstract}
\maketitle

In 1922 Albert Einstein, the physicist, met in Paris Henri Bergson, the philosopher. The two giants debated publicly about time and Einstein concluded with his famous statement: "There is no such thing as the time of the philosopher". Around the same time, and equally dramatically, mathematicians were debating how to describe the continuum. The famous German mathematician David Hilbert was promoting formalized mathematics, in which every real number with its infinite series of digits is a completed object. On the other side the Dutch mathematician, Luitzen Egbertus Brouwer, was defending the view that points on the line should be represented as a never-ending process that develops in time, a view known as intuitionistic mathematics. Although Brouwer was backed-up by a few well-known figures, like Weyl \cite{Weyl} and G\"odel \cite{Godel}, Hilbert and his supporters clearly won that second debate. Hence, time was expulsed from mathematics and mathematical objects came to be seen as existing in some idealized Platonistic world.

These two debates had a huge impact on physics. Mathematics is the language of physics and Platonistic mathematics makes it difficult to talk about time. Hence, the sense of flow of time was also expulsed from physics: all events are the ineluctable consequence of some "quantum fluctuations" that happened at the origin of time - at the big-bang. Accordingly, in today’s physics there is no "creative time" and no "now". This had dramatic consequences, in particular when one remembers that physics is not only about technologies and abstract theories, but also about stories that tell how nature functions. Time is an indispensable ingredient in all human narratives. As Yuval Dolev emphasized, "To think of an event is to think of something in time. [...] Tense and passage are not removable from how we think and speak of events" \cite{DolevSilence}. So, it seems that physics should give up telling stories and concentrate on more and more abstract theories. But is this really necessary? Isn’t physics in danger of coming to a hold when faced with claims like "time is an illusion" \cite{BarbourEndTime}. Shouldn’t one adapt Rabelais’ famous sentence: "Science without time is only ruin of intelligibility"?

Independently of these debates, in \cite{NGrealNb} I argued that a finite volume of space can’t contain more than a finite amount of information and concluded that physically relevant numbers can’t contain infinite information. Then, in a workshop in Jerusalem, I discovered thanks to Prof. Carl Posy intuitionistic mathematics which comes surprisingly close to my – and I bet many physicists – intuition about the continuum \cite{Posy76, PosyBook}. In this comment I like to share my excitement about this finding. 

I believe physicists should revisit these debates. The mathematical language that scientists use makes it easy or difficult to formulate some concepts, like the passage of time. 
Bergson never agreed with Einstein’s statement and Einstein himself felt uncomfortable with his beloved physics lacking the concept of "now", though, admittedly, he didn’t see any way to incorporate it and thus concluded that one has to live with this state of affairs. Also Hilbert felt uncomfortable by the infinities that his beloved axiomatized mathematics introduced, stating clearly that physics should never incorporate actual infinities \cite{HilbertInfinite, NPcommentEllis}.

In order to illustrate the tension physics faces, let us consider classical mechanics, a field very well understood, including its limiting cases, relativity and quantum theory, and a field with an enormous explanatory power. Although, classical mechanics is usually illustrated with clocks and other cog-wheels, almost all classical dynamical systems are of a very different nature: they are chaotic. Because of the hyper-sensitivity on initial conditions, the future of typical classical systems depends on far down the series of digits of the initial conditions. A simple but relevant example of typical classical dynamical systems can be modelled as follows. Assume the state-space reduces to the unit interval and consider discrete time-steps. The state is thus described by a number formed by an infinity of digits bn that follow zero. At each time-step the digits are shifted by one place to the left, with the first digit dropping out:

\beqa
X_t &=&   0. b_1 b_2 b_3 b_4 ... b_n ...		\nonumber\\
X_{t+1} &=& 0. b_2 b_3 b_4 ... b_n ...
\eeqa
							 					
After n time-steps, the most relevant digit is the nth digit of the initial condition. This most relevant digit could, for example, represent today’s weather (let’s say 0 for pouring rain and 9 for bright sun). If one assumes all digits of the initial condition faithfully represent physical reality, then, according to our simple model, the weather of the entire future is already fully determined. This illustrates the absence of any "creative time"; nothing really happens, everything is determined by the initial condition and the deterministic evolution equations. At least, this is the consequence of representing the points on the line between 0 and 1 by classical Platonistic mathematics. 

But is such a representation truly faithful and necessary? My experience is that most physicists reject the faithfulness of such a representation, but admit – often with regret – that they don’t see how one could do otherwise. For example, Max Born, one of the fathers of quantum theory, stressed that ``Statements like `a quantity x has a completely definite value' (expressed by a real number and represented by a point in the mathematical continuum) seem to me to have no physical meaning" \cite{Born} (see also \cite{DowekRealNb13} for the view of a computer scientist). 

This is where intuitionistic mathematics can help.
In intuitionistic mathematics numbers are processes that develop in time; at each moment of time, there is only finite information. One way to understand this unusual claim goes as follows. Assume nature has the power to produce random numbers. One may think of a quantum random number generator. That would do, but here it is preferable not to think of a human-made randomness source, but rather as a power of nature: nature is intrinsically and fundamentally indeterministic. 
Now, this source of randomness feeds the digits of typical real numbers, as illustrated in figure 1. Let me emphasize that the digits of all typical real numbers are truly random - as random as the outcomes of quantum measurements, as has been nicely emphasized by, for instance Gregory Chaitin \cite{Chaitin1, Chaitin2}. Moreover, typical real numbers contain infinite information, allowing one, for example, to code in a single number the answers to all questions one may formulate in any human language, as noticed by Emile Borel \cite{Borel}. 

So, we face a choice. Either all digits of the initial conditions are assumed to be determined from the first moment, leading to timeless physics; or these digits are initially truly indeterminate and physics includes events that truly happen as time passes. Notice that in both perspectives chaotic systems would exhibit randomness. In the first case, from the point of view based of classical Platonistic mathematic, all the randomness is encoded in the initial condition. In the second case, randomness happens as time passes - as described by intuitionistic mathematics, were the dependence on time is essential \cite{StandfordEncyclodediaIntuitionism}. One may object that intuitionism doesn’t derive indeterminism, but assumes it from the start. Correct. Likewise, classical mathematics assumes actual infinite information from the start.

The two views cannot be distinguished empirically. One can always claim that instead of God playing dice every time a random outcome happens, God played all the dice at once before the big-bang and encoded all results in the Universe’s initial condition. Despite the empirical equivalence of the two views, they present us with very different pictures of our world. Somehow, the real numbers are the hidden variables of classical mechanics \cite{NGHiddenReals}.

\begin{figure}[h]
\includegraphics[width=6cm]{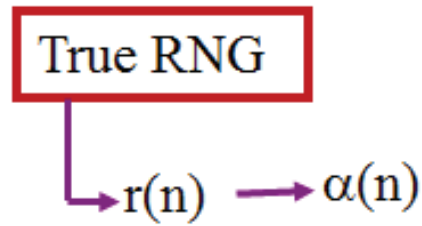}
\caption{\it A way to understand the continuum in intuitionistic mathematics, well suited to physicists, assumes that nature has the power to produce true randomness, here illustrated as a "True Random Number Generator" that outputs a digit $r(n)$ at each time step $n$. Then, at each time-step $n$, a rational number $\alpha(n)$ is computed by a function $f$ : $\alpha(n)=f\big(\alpha(n-1), n, r(1),…,r(n)\big)$. Different functions $f$ define different classes of series $\alpha(n)$. The series $\alpha(n)$ is assumed to converge, however, at any time, only finite information about the series exists, in accordance with the basic idea that the random number generator is a genuine endless process that develops in times. A first simple example assumes that the function $f$ merely adds the random digit $r(n)$ as the $n$th digit of $\alpha$, i.e. $\alpha(n) = \alpha(n-1)+r(n)\cdot 10^{-n} = 0.r(1)r(2)r(3)...r(n)$. Usually, it is assumed that $\alpha(0)$ is a given initial rational number, but that is not essential. Note that if the random number generator is actually a pseudo-random number generator, then $\alpha(n)$ converges to a computable number. In other examples, the digits of $\alpha(n)$ are correlated, e.g. new digits of $\alpha(n)$ depend on the previous $k$ random numbers $r(j)$. By choosing suitable functions $f$, an infinity of classes of series can be defined. Historically, Brouwer, the father of intuitionism, did not use random number generators, but mathematical objects he named choice sequences, where the choices were made by an idealized mathematician \cite{Brouwer1948}, and what I name "classes" were termed "spreads" by Brouwer.}
\label{Fig1}
\end{figure}

Intuitionism goes way beyond the description of the continuum. Physicists often have the intuition that the present is thick, that the present is not of measure zero. This corresponds naturally to the intuitionistic concept of the viscous continuum: the continuum cannot be sharply cut in two. When trying to cut it, something sticks on the knife. This aspect of intuitionistic mathematics is closely related to the law of the excluded middle which does not hold in intuitionistic logic. A proposition P could be neither true nor false. This is very difficult to swallow for today’s scientists who have been fed above reason by classical mathematics. But think of a proposition about the future, e.g., "It will be raining in exactly one year time from now at Piccadilly Circus". If one believes in determinism, then this proposition is either true or false, though it might be impossible in practice to know which alternative holds. But if one believes that the future is open, then it is not predetermined that it will rain, hence the proposition is not true, and it is not predetermined that it will not rain, thus the proposition is also not false. There are similar examples using choice sequences $\alpha(n)$. Assume the random numbers are not digits, but binary $r(n)=\pm1$, and let $\alpha(n)=1/2+r(n)\cdot 10^{-n}$. This goes on until, by chance, $n/2$ consecutive $r(j)$’s take the same value (and $n$ is even and larger than 2), after which the series terminates and all future $\alpha(n)$ remain constant. Since at each time-step the probability that the series terminates decreases exponentially, there is a finite probability that $\alpha(n)$ oscillates forever between below and above 1/2. Accordingly, as long as the series did not terminate the proposition $\alpha<1/2$ is neither true nor false. Hence the continuum is viscous \cite{Posy76}: it can’t be sharply cut in two, above and below 1/2. In these two examples, the law of the excluded middle holds only if one assumes a look from the "end of time", i.e. a God’s eye view. But at finite times, intuitionism states that the law of the excluded middle is not necessary, that time truly passes and the future is open. Looking at the law of the excluded middle in this way makes its absence in intuitionistic mathematics easily acceptable and renders the present naturally thick.

\begin{table}[ht]
	\centering
		\begin{tabular}
			{|c|c|}
			\hline
			\large
			Indeterministic & \large Intuitionistic \\ \large Physics & \large Mathematics\\
			\hline
Past, present and future	& Digits of real numbers \\
 are {\bf not} all given at once & are {\bf not} all given at once\\
			\hline
			Time passes	& Numbers are processes \\
			\hline
			Indeterminism &	Numbers contain finite information\\
			\hline
The present is thick	& The continuum is viscous \\
			\hline
Experiencing	& Intuitionism\\
			\hline
Becoming &	Choice sequences\\
			\hline
The future is open &	No law of the excluded middle \\&(a proposition about the future \\&can be neither true nor false)\\
			\hline
		\end{tabular}
\caption{\it Intuitionistic mathematics nicely allows one to express concepts that appear natural in indeterministic physics.}
\end{table}

Let me emphasize that classical formal mathematics and real numbers are marvelous tools that should not be abandoned. However, their practical use should not blind the physicists; after all, their use does not force us to believe that "real numbers are really real" \cite{NGrealNb}. In other words, one should not confuse the epistemological usefulness of classical mathematics with the ontology, which might well be better described by intuitionistic mathematics. Furthermore, in practice one never uses real numbers with all its infinitely many digits. Think, for example, of computer simulations that necessarily at each time only consider finite information numbers. Scientists working on weather and climate physics explicitly use finite-truncated numbers and stochastic remainders \cite{PalmerStochClimateModel}.
All those who speak more than one language know that some concepts are easier to express in one language than in another. I believe the same is true of the language of physics. Which mathematics we adopt when "talking physics" will influence the possibility of indeterminism in physics, see table 1. I believe that the notion of a deterministic and timeless world does not arises from the huge empirical success of physics, but from taking Platonistic mathematics to be the only language for physics. Physics can be as successful if build on intuitionistic mathematics, even if this breaks its wedding to determinism.
This original take on the mathematical language used in physics provides an interesting and, in my opinion, very positive perspective: contrary to usual expectations, I bet that the next physics theory will not be even more abstract than quantum field theory, but might well be closer to human experience.

\small
\section*{Acknowledgment} This work was supported by the Swiss NCCR SwissMap.  
It appeared as a comment to Nature Physics in January 2020 \cite{NPc}\\

\end{document}